%% file: LT.tex
\documentclass[PP]{AEA}
\usepackage{latexsym,amssymb}
\usepackage{graphicx}
\usepackage{multicol}
\usepackage{harvard}
\usepackage{url}
\usepackage{bbm}
\usepackage{caption}
\usepackage{enumitem}
\usepackage{color}
\def\GG{\mathcal G}

\def\NN{\mathcal N}

\def\11{\mathbbm{1}}

\begin{document}
\bibliographystyle{aea} 
\title{Ranking and Selection from Pairwise Comparisons:\\
Empirical Bayes Methods for Citation Analysis}
\shortTitle{Pairwise Ranking and Selection} 
\author{Jiaying Gu and Roger Koenker\thanks{Gu: University of Toronto, Toronto, Ontario, M5S 3G7, Canada, jiaying.gu@utoronto.ca.  Koenker: University College London, London, WC1H 0AX, UK, r.koenker@ucl.ac.uk.  Jiaying Gu acknowledges financial support from Social Sciences and Humanities Research Council of Canada under Insight Grant 504477.  
The authors wish to thank Steve Stigler for comments.}}
\maketitle

    Competitive play among pairs of individuals or teams naturally leads to attempts to construct
    ratings and rankings of players.  No where is this more evident than in sports competition, but
    there are manifestations in almost every field of endeavor.  The question, ``Who are the best?''
    demands an answer, and where there is demand there is bound to be a supply.\footnote{For a broader
    perspective, related recent work on ranking and selection problems not confined to the pairwise comparison 
    setting includes: \cite{MRSW}, \cite{GH}, \cite{AKM} and \cite{GK}.}

    The most common statistical model for rating and rankings based on paired comparison data is 
    the \cite{BT} model anticipated by \cite{Zermelo} as a proposed method for rating
    chess players.  Given $p+1$ players of unknown ``abilities,'' $\alpha_0, \cdots , \alpha_p$, it is
    postulated that the probability of player $i$ defeating player $j$ is given by,
    \[
    \pi_{ij} = \alpha_i / ( \alpha_i + \alpha_j),
    \]
    and with a sufficiently rich history of competition player ratings can be estimated by maximum
    likelihood.

    It is convenient to reparameterize abilities so $\theta_i = \log \alpha_i$ and $\pi_{ij}$, becomes,
\[
\pi_{ij} = \frac{1}{1 + \exp(-(\theta_i - \theta_j))}.
\]
The (logistic) log likelihood for $n$ binary outcomes, $y_1, y_2, \dots , y_n$, can be written
with $h_\theta (x_k) = 1/(1 + \exp(-\theta^\top x_k))$,  as,
    \begin{align*}
\ell (\theta | y) = \sum_{k=1}^n y_k & \log(h_\theta (x_k))\\
& + (1 - y_k) \log(1 - h_\theta (x_k)) 
\end{align*}
where for match $k$ between $i$ and $j$, $x_k$ is an $p$ vector  with $i$th element
1, and $j$th element -1, and other elements 0.  Without loss of generality, we can set $\theta_0 = 0$.
When there are multiple meetings between a given pair, $(i,j)$, binary outcomes can be aggregated
to write the likelihood in terms of the corresponding binomial outcomes.  In this case the log likelihood
can be expressed in terms of the $\alpha_i$'s as,
\[
\ell(\alpha | y) = K + \sum_{i=1}^p  w_i \log \alpha_i - \sum_{i < j} n_{ij} \log (\alpha_i + \alpha_j),
\]
where $w_i$ denotes the total number of ``wins'' of player $i$ against all other competitors, and $n_{ij}$ 
is the total number of matches between players $i$ and $j$.  From this it may be concluded that the
This observation has played an important role in recent literature on the analysis of paired comparisons
in machine learning where A/B experiments constitute an important data structure, see for example
\cite{Shah}.  However, this sufficiency
result holds if and only if the $n_{ij}$ are uninformative about the ratings.  This is plausible  if matching
is generated at random or by some balanced tournament design, but certainly cannot be expected to
hold if, for example, players are more likely to be matched with players of similar ability.  In such cases
one must resort to the MLE which offers protection against such variability in ``strength of schedule.''
Just counting wins is insufficient in designs like this when the $n_{ij}$ are informative too.

Since the number of players, $p$, can be large relative to the number of matches $n$, some regularization
of the MLE estimates may be beneficial.  We compare two such regularization schemes one employing a variant
of the familiar $\ell_1$ penalty that encourages grouping of the estimated abilities thereby reducing their
effective dimensionality, and another empirical Bayes approach that treats the ratings as if they arise from
a nonparametric Gaussian mixture model.  Comparisons with vanilla MLE ranking and two variants
of Borda score ranking illustrates advantages of the regularization methods.

\section{Grouped Lasso for Ranking}

In some ranking problems it is plausible that there are only a few equivalence classes of ability, groups
of players that are indistinguishable in terms of proficiency.  For the Bradley-Terry model this suggests
penalization of the MLE that would shrink pairwise differences of the rating parameters toward zero. This
is conveniently accomplished by solving,
\[
\min \{ - \ell (\alpha | y) + \lambda \| D \alpha \|_1 \},
\]
where $\| D \alpha \|_1 = \sum_{i < j} |\alpha_i - \alpha_j |$.  This penalty has been proposed by 
\cite{Bach} and employed by \cite{Firth} to study journal rating and ranking.  As $\lambda$
is increased the $\alpha$ parameters are pulled together into fewer and fewer groups.  This is somewhat
analogous to total variation regularization for nonparametric smoothing where piecewise linear fitting
selects only a few places at which the derivative of the estimated function jumps.  The problem is 
convex and thus efficiently solved by interior point methods; our preferred solver is Mosek,
\cite{Mosek}. 

\section{Empirical Bayes Posterior Mean Ranking}

Another approach to regularization is to treat the unconstrained logistic estimates, $\hat \theta$,
as approximately independent draws from a Gaussian sequence model.  When the problem design is unbalanced
so the number of  matched pairs are not equal, the MLE point estimates will have different precision and
off-diagonal elements of their covariance matrix are also more heterogeneous.  Initially, we will ignore
the latter aspect and treat the estimated maximum likelihood rating parameters as a sample from a Gaussian
sequence model with heterogeneous scale parameters.  

Since the pairs, $(\hat \theta , \hat \sigma) = 
\{(\hat \theta_i , \hat \sigma_i), \quad i = 1, \cdots ,p \}$ can be  relatively high dimensional compared to 
the number of matches, $n$, another regularization strategy is to treat the $\hat \theta_i$'s as if the arose
from the mixture density,
\[
f (\theta | \hat \theta_i , \hat \sigma_i ) = \int \varphi_{\hat \sigma_i} (\theta - \hat \theta_i ) 
dG(\theta),
\]
where $\varphi_{\sigma}$ denotes of the Gaussian density with mean zero and variance, $\sigma^2$.
The mixing distribution $G$ can be estimated by maximum likelihood as suggested by \cite{R50},
\cite{KW}, \cite{Laird}, \cite{JZ}, and \cite{KM}, by solving,
\begin{align*}
\min_{G \in \GG} \{ - \sum_{i=1}^p & \log f_G (\hat \theta_i) \; |\\ 
& f_G( \hat \theta_i ) = \int \varphi_{ \hat \sigma_i} ( \hat \theta_i - t) dG(t) \}.
\end{align*}
This problem is again convex and can be efficiently solved with the interior point implementation of Mosek.  
Given a $\hat G$ we can compute a posterior means for each $\theta_i$ and from these ratings, 
rankings may be constructed.

\section{Empirical Bayes Posterior Ranks}
Alternatively, we can use estimated ratings to construct ranks based on pairwise comparisons of the ratings,
\[
R_i = \sum_{j \neq i} \11 \{\alpha_i \geq \alpha_j\}.
\]
When the $\alpha$'s are Gaussian \cite{LL} have considered the quadratic loss, $\sum_{i=1}^n (\hat R_i - R_i)^2$,
under which the Bayes rule is the posterior mean rank.  We can approximate this by,
\begin{align*} 
\hat R_i & = \sum_{j \neq i} \mathbb{P}(\alpha_i \geq \alpha_j \; | \; \hat \alpha_1, \ldots , \hat \alpha_n ) \\
& = \sum_{j \neq i} \frac{\int_{\alpha_i \geq \alpha_j} \varphi_{ij}((\hat \alpha_i, \hat \alpha_j)) 
d \hat G(\alpha_i) d \hat G(\alpha_j)}{ \int \varphi_{ij}((\hat \alpha_i, \hat \alpha_j))d \hat G(\alpha_i) d \hat G(\alpha_j)}
\end{align*} 
where $\varphi_{ij}(z)$ is a bivariate Gaussian density with mean, $\mu = (\alpha_i, \alpha_j)$ 
and covariance matrix, $\Sigma(i,j)$, estimated from the Hessian of the MLE.
The mixing distribution, $G$ is estimated by the maximum likelihood procedure described in the previous section. 

\section{A Simulation Exercise}

We now compare performance of seven rating procedures intended to rank latent abilities:
\begin{description}
	\item[MLE] Logistic MLE
	\item[KWPM] Posterior Mean Ratings
	\item[KWPMs] Smoothed Posterior Means
	\item[KWPR] Posterior Mean Ranks
	\item[RMLE] Group Lasso MLE
	\item[B] Borda Scores
	\item[WB] Weighted Borda Scores
\end{description}
Performance is evaluated by computing the mean of the Kendall rank correlation between the estimated true abilities
and the estimated ratings for each procedure for 100 replications.  Comparisons based Spearman's rank correlation
yield very similar conclusions, see \cite{DG}.  In all the experimental settings there are 100
players. There are three other design dimensions of the experiment:  
\begin{itemize}
    \item  Distribution of Latent Abilities
    \begin{description}
	\item $\alpha \sim e^Z + 2, \; Z \sim \NN (0,1)$
	\item $\alpha \sim 0.8 \delta_4 + 0.2 \delta_8 + Z/3$
    \end{description}
    \item Matching Design
    \begin{description}
	\item[RS] Random Pairing
	\item[LS] Similar Ability Pairing
    \end{description}
    \item Sample Size
\end{itemize}

True abilities are either lognormal or a mixture of two Diracs contaminated by Gaussian noise.
Pairings are generated completely at random or by pairing players of similar ability.
In the random matching case Borda scores can be expected to perform well, however when pairings
are based on similar abilities Borda scores, which ignore ``strength of schedule,'' falter.  

\begin{figure} \includegraphics[width=\textwidth]{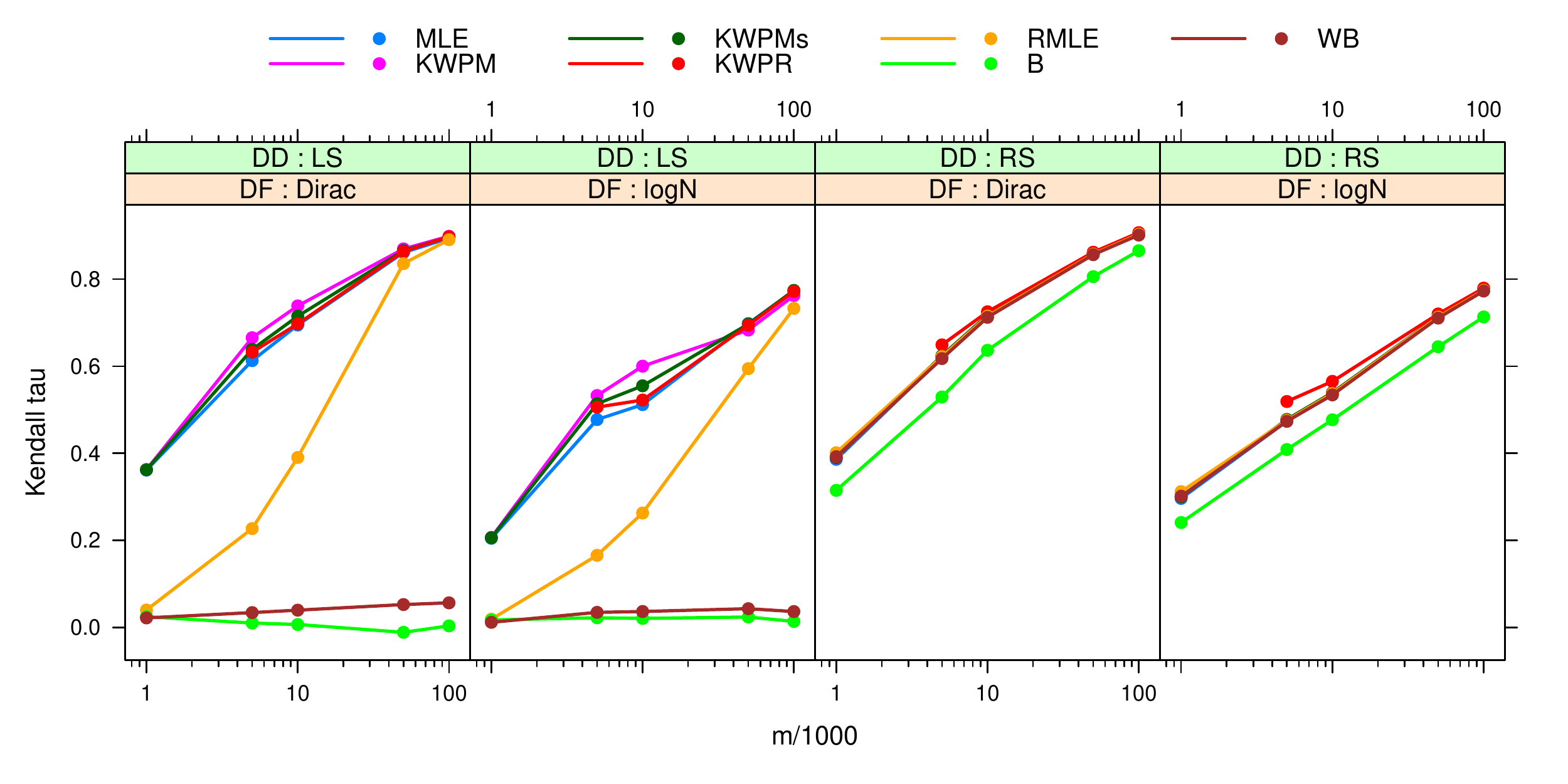}
   \caption{Simulation Comparison of Several Ranking Methods} \label{fig.sim1-4}
\end{figure}

We consider five sample sizes $\{1000, 5000, 10000, 50000, 100000 \}$.  With 100 players, 1000 pairings 
implies teams play on average only 10 matches, so the MLE is quite noisy and shrinkage isn't very
effective.  At the other extreme with 100,000 pairings, so teams on average have 1000 matches the
MLE is quite accurate and there is also little room for improvement from shrinkage.  For intermediate
sample sizes we see considerable benefit from some of the shrinkage methods.  The noisy  Dirac
ability mixture is particularly favorable for the unsmoothed Kiefer-Wolfowitz method, but it performs
well when the ability distribution is lognormal as well.  Performance of the group lasso method is
rather disappointing, but this might be attributed to poor choice of the smoothing parameter $\lambda$. 
Both the smoothed posterior means and the posterior mean ranks rely on a smoothed version of the 
Kiefer-Wolfowitz $\hat G$, so better tuning of their bandwidth choices might improve their performance.

Figure \ref{fig.sim1-4} reports results for 100 players and 100 replications.
When players of similar ability are more likely to meet, (DD:LS), the Borda methods
of estimating ranks fails miserably, and there is a modest improvement from regularization
over the performance of the MLE in this setting for both the noisy Dirac and  lognormal
ability distributions.  With random matching Borda performs credibly and other procedures
are almost indistinguishable. 
The grouped lasso is also disappointing, but this may be attributable to poor $\lambda$ selection.

\section{The Stigler Model of Journal Influence}

\cite{Stigler} considers a model of journal influence based on pairwise citation counts.
Citations in journal $j$ of papers appearing in journal $i$ are viewed as a measure of the
influence of journal $i$ on journal $j$.  Ratings of journal influence are formulated as a
Bradley-Terry model and rankings can be evaluated from estimated ratings.  We consider pairwise 
citation counts  for 86 journal spanning the econometrics/statistics fields over the decade
2010-2019 as reported in  Clarivate Journal Citation Reports.  Self citations, that is citations
by a journal to papers that appeared in the same journal, while perhaps interesting, are ignored
in the subsequent analysis.

{\centering
    \includegraphics[width=\columnwidth]{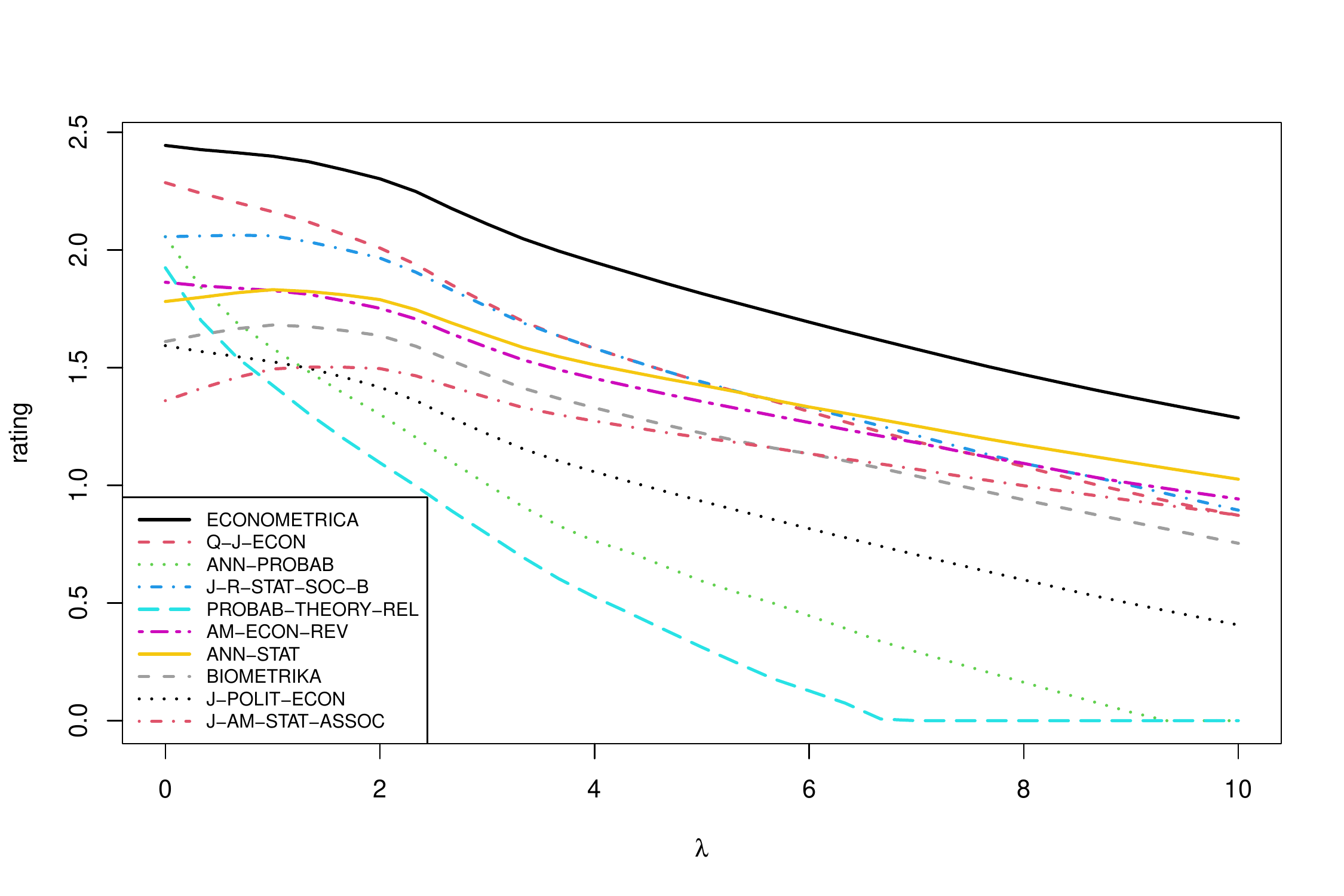}
    \captionof{figure}{Grouped Lasso Shrinkage Plot} \label{fig.Rlasso10}
    }

{\centering
    \includegraphics[width=\linewidth]{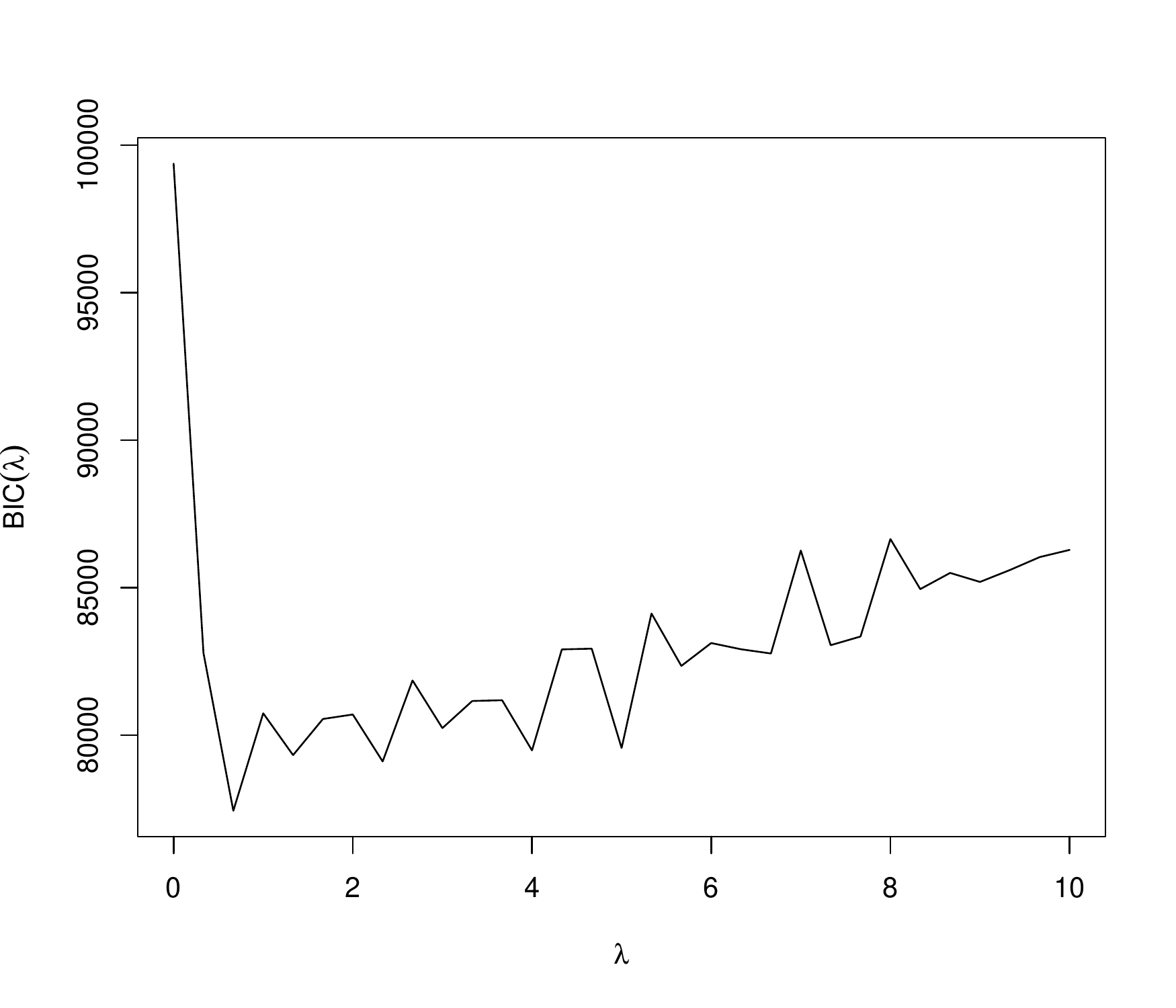}\\
    \captionof{figure}{BIC $\lambda$-Selection Plot} \label{fig.BIC}
    }

{\scriptsize{
\input RTop.tex
}}

Figure \ref{fig.Rlasso10} illustrates the shrinkage effects of the grouped lasso procedure for the top
10 journals according to the maximum likelihood estimates.  \emph{Econometrica}  is the most highly rated
journal and remains so over the whole range of $\lambda \in [0,10]$.  The \emph{Quarterly Journal of Economics}
appears to be a strong second place finisher, but careful inspection of the plot reveals that the MLE 
very slightly favors the \emph{Annals of Probability}.  It may seem surprising that both probability
journals are so highly rated, but a little reflection reveals that although they are rarely cited by
the other journals they almost never cite these journals; this is the hallmark of an influential
journal.  Once shrinkage is applied the probability journal fade into obscurity and more mainstream
journals acquire more prominence.  This raises the inevitable question:  How should one choose $\lambda$?
This is always controversial.  If we adopt the \cite{Schwarz} BIC criterion that penalizes parametric
dimension, in this case the number of estimated groups, which decreases with $\lambda$, we get the plot
appearing in Figure \ref{fig.BIC} showing that only a very modest amount of shrinkage is desirable.

Since the grouped lasso doesn't alter the ranking of the top journals, we turn our attention to
our empirical Bayes procedure.  We consider ranking based on the posterior mean of the rating parameters.
As noted earlier this procedure involves no tuning parameter selection.  If the maximum likelihood rating
estimates were based on a balanced design and consequently had homogeneous standard errors, the posterior
mean Bayes rule would produce the same ranking as the MLE.  The Bayes rule must be monotone it such 
circumstances.  Our citation data is far from balanced, so there may be potentially different rankings.
Table \ref{RTop} reports ranking for our five procedures

As anticipated the grouped lasso regularization doesn't alter the MLE ranking, and the Kiefer-Wolfowitz
posterior mean ranking only flips the \emph{Annals of Probability} and \emph{JRSS(B)}.  
The Borda rankings deprecate both probability journals and elevate \emph{JASA} and \emph{JRSS(B)}.

\section{Conclusion}

We have considered a very special but commonly employed model for inferring latent abilities from pairwise
comparison data.  Although there is a recent wave of theoretical support for simple estimators like 
Borda scores, we have seen that good performance of such measures depends critically on balanced design
assumptions that may not be plausible in applications.  In contrast, the classical logistic MLE
underpinning the Bradley Terry model offers a more robust alternative approach.  When the number of 
``players'' is large there are opportunities to improve upon the performance of the MLE by various forms
of shrinkage including a variant of the group lasso and several empirical Bayes methods based on the
Kiefer-Wolfowitz nonparametric MLE.  As we have argued elsewhere decision making for rating, ranking
and selection are inextricably intertwined and rules for ranking can be adapted to rules for selection
that bring into play control of false discovery rates and related issues.

There are several intriguing directions for future investigation.  Optimal design of pairwise matching
schemes particularly with dynamic selection seems challenging, but potentially very rewarding.  Connections to
network formation models and their analysis may be fruitful as exemplified by the reliance on results
of \cite{SY} in the work of \cite{Graham}.  Finally, there is considerable scope for moving away from the
stringent assumptions of the Bradley-Terry model to consider more flexible formulations.

\bibliography{LTables}
\end{document}

%% file: RTop.tex
\begin{table}[!tbp]
\begin{center}
\begin{tabular}{lrrrrr}
\hline\hline
\multicolumn{1}{l}{}&\multicolumn{1}{c}{MLE}&\multicolumn{1}{c}{RMLE}&\multicolumn{1}{c}{KWPM}&\multicolumn{1}{c}{Borda}&\multicolumn{1}{c}{WBorda}\tabularnewline
\hline
ECONOMETRICA&$ 1$&$ 1$&$ 1$&$ 4$&$ 1$\tabularnewline
Q-J-ECON&$ 2$&$ 2$&$ 2$&$ 5$&$ 3$\tabularnewline
ANN-PROBAB&$ 3$&$ 3$&$ 4$&$13$&$ 7$\tabularnewline
J-R-STAT-SOC-B&$ 4$&$ 4$&$ 3$&$ 7$&$ 2$\tabularnewline
PROBAB-THEORY-REL&$ 5$&$ 5$&$ 5$&$19$&$ 9$\tabularnewline
AM-ECON-REV&$ 6$&$ 6$&$ 6$&$ 1$&$ 6$\tabularnewline
ANN-STAT&$ 7$&$ 7$&$ 7$&$ 3$&$ 4$\tabularnewline
BIOMETRIKA&$ 8$&$ 8$&$ 8$&$ 6$&$ 5$\tabularnewline
J-POLIT-ECON&$ 9$&$ 9$&$ 9$&$ 9$&$10$\tabularnewline
J-AM-STAT-ASSOC&$10$&$10$&$10$&$ 2$&$ 8$\tabularnewline
\hline
\end{tabular}
\caption{Comparison of Top Ten Journal Influence Rankings for Five Methods\label{RTop}}\end{center}
\end{table}